\documentclass{sig-alternate}

\usepackage{subfigure}
\usepackage{graphicx}

\usepackage{url}
\usepackage{amsmath}
\usepackage{multicol}
\usepackage{fancyvrb}
\usepackage{float}
\usepackage{array}
\usepackage{multirow}
\usepackage{tabularx}
\usepackage{colortbl}
\usepackage{hhline}
\usepackage{textcomp}

\usepackage[T1]{fontenc}
\usepackage[utf8]{inputenc}
\usepackage{lmodern}

\usepackage[linesnumbered]{algorithm2e}

\begin{document}

%\title{Identifying Deferred Representations for Archiving Using Both Heritrix and PhantomJS}
%\title{How to Crawl JavaScript Web Pages: Understanding their Impact on Web-Scale Archiving}
\title{Adapting the Hypercube Model to Archive Deferred Representations and Their Descendants}

\numberofauthors{1}

\author{
% 1st. author
\alignauthor
Justin F. Brunelle, Michele C. Weigle, and Michael L. Nelson\\
       \affaddr{Old Dominion University}\\
       \affaddr{Department of Computer Science}\\
       \affaddr{Norfolk, Virginia, 23508}\\
       \email{\{jbrunelle, mweigle, mln\}@cs.odu.edu}
}

\maketitle
\begin{abstract}
The web is today’s primary publication medium, making
web archiving an important activity for historical and analytical
purposes. Web pages are increasingly interactive,
resulting in pages that are increasingly difficult to archive.
Client-side technologies (e.g., JavaScript) enable interactions
that can potentially change the client-side state of a representation.
We refer to representations that load embedded
resources via JavaScript as \emph{deferred representations}. It is
difficult to archive all of the resources in deferred representations
and the result is archives with web pages that
are either incomplete or that erroneously load embedded resources
from the live web.

We propose a method of discovering and crawling deferred
representations and their \emph{descendants} (representation states
that are only reachable through client-side events). We
adapt the Dincturk et al. Hypercube model to construct a
model for archiving descendants, and we measure the number
of descendants and requisite embedded resources discovered
in a proof-of-concept crawl. Our approach identified an
average of 38.5 descendants per seed URI crawled, 70.9\% of
which are reached through an onclick event. This approach
also added 15.6 times more embedded resources than Heritrix
to the crawl frontier, but at a rate that was 38.9 times
slower than simply using Heritrix. We show that our dataset
has two levels of descendants. We conclude with proposed
crawl policies and an analysis of the storage requirements
for archiving descendants.

\end{abstract}

% A category with the (minimum) three required fields



\section{Introduction}
\label{intro}

As the web grows as the primary medium for publication, communication, and other services, so grows the importance of preserving the web (as evidenced by recent articles in The New Yorker \cite{newyorker}, NPR \cite{npr}, and \emph{The Atlantic} \cite{lafrance}). Web resources are ephemeral, existing in the perpetual \emph{now}; important historical events frequently disappear from the web without being preserved or recorded. We may miss pages because we are not aware they should be saved or because the pages themselves are hard to archive. 

\begin{figure}[h!]
    \includegraphics[width=0.4\textwidth]{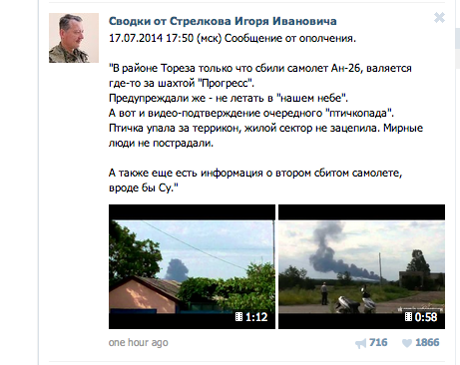}
  \caption{A screenshot of the Ukrainian Separatists' announcement.}
\label{shotdown}
\end{figure}

On July 17, 2015, Ukrainian separatists announced via social media\footnote{VKonkakte, \url{https://vk.com/}, is a Russian social media site.}, with video evidence, that they shot down a military cargo plane in Ukrainian airspace (Figure \ref{shotdown}). However, the downed plane was actually the commercial Malaysian Airlines Flight 17 (MH17). The Ukrainian separatists removed from social media their claim of shooting down what we now know was a non-military passenger plane. The Internet Archive \cite{iawebarchive}, using the Heritrix web crawler \cite{heritrix, Sigurosson:Incremental-Heritrix}, was crawling and archiving the social media site twice daily and archived the claimed credit for downing the aircraft; this is now the only definitive evidence that Ukrainian separatists shot down MH17 \cite{csm}. This is an example of the importance of high-fidelity web archiving to record history and establish evidence of information published on the web.

%For example, the SOPA protest by Wikipedia has not been accurately preserved by the web archives.

\begin{figure*}
  \begin{center}
    \subfigure[A screenshot of the Wikipedia SOPA protest taken during the protest in 2012.]{\label{wiki1}\includegraphics[width=0.4\textwidth]{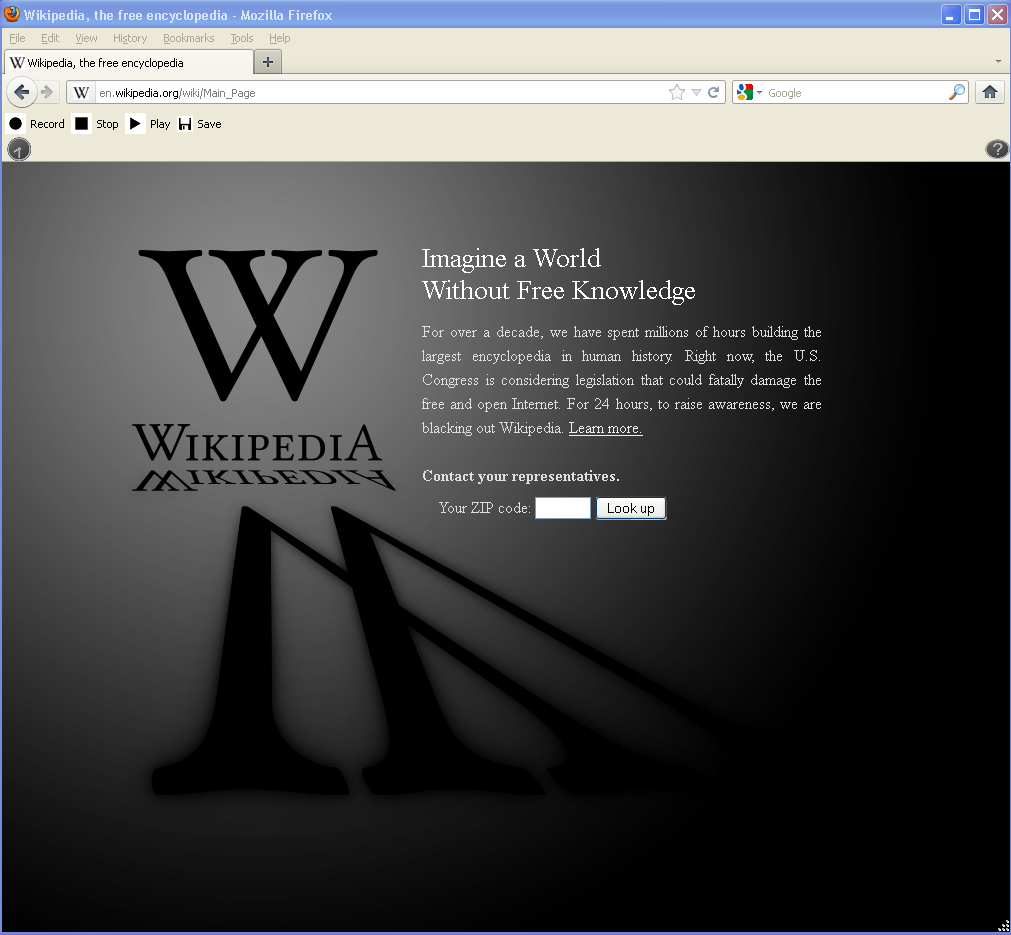}}\qquad
    \subfigure[The Internet Archive memento of the SOPA blackout does not include the JavaScript-loaded splash page.]{\label{wikisopa2}\includegraphics[width=0.52\textwidth]{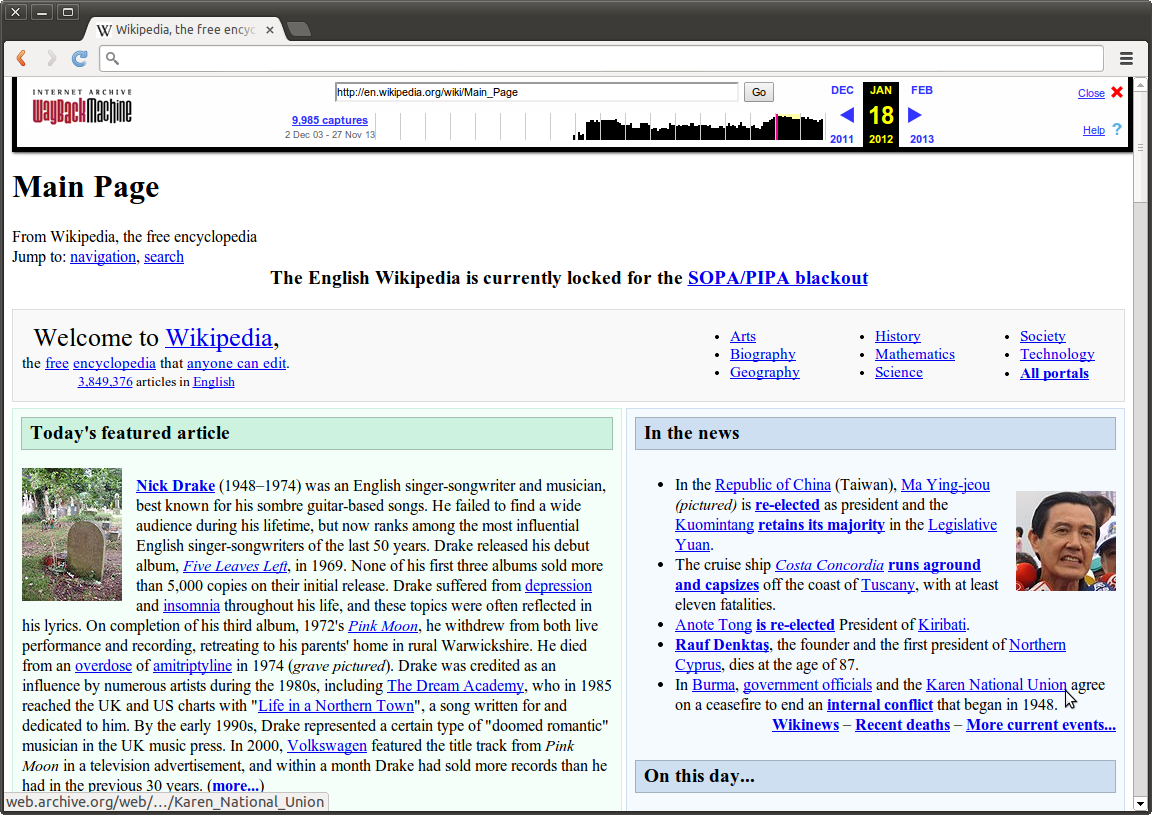}}
  \end{center}
  \caption{Screenshots of the Wikipedia blackout in protest of SOPA live and in the Internet Archive.}
  \label{liveWiki}
\end{figure*}

However, not all historical events are archived as fortuitously as the MH17 example. In an attempt to limit online piracy and theft of intellectual property, the U.S. Government proposed the widely unpopular Stop Online Piracy Act (SOPA) \cite{sopawiki}. While the attempted passing of SOPA may be a mere footnote in history, the overwhelming protest in response is significant. On January 18, 2012, many prominent websites organized a world-wide blackout in protest of SOPA \cite{sopapost, sopaabc}. Wikipedia blacked out their site by using JavaScript to load a ``splash page'' that prevented access to Wikipedia's content (Figure \ref{wiki1}).

The Internet Archive, using Heritrix, archived the Wikipedia site during the protest. However, the archived January 18, 2012 page\footnote{\url{http://wayback.archive-it.org/all/20120118184432/http://en.wikipedia.org/wiki/Main_Page}}, as replayed in the Wayback Machine \cite{waybackarchives2}, does not include the splash page (Figure \ref{wikisopa2}) \cite{brunelleSopa}. Because archival crawlers such as Heritrix are not able to execute JavaScript, they neither discovered nor archived the splash page. Wikipedia's protest as it appeared on January 18, 2012 has been lost from the archives and, without human efforts, would be potentially lost from human history.

The SOPA protest, like MH17, is an example of an important historical event. Unlike the MH17 example (which establishes our need to archive with high fidelity), the SOPA example is not well represented in the archives. In this work, we present a method to improve the fidelity of JavaScript-dependent archival copies. %Specifically, we show that archival crawlers can use PhantomJS to interact two levels deep into a representation, uncovering 15.6 times more embedded resources (70.9\% of which are available via onclick events).
Specifically, we show that archival crawlers that use PhantomJS can interact just two levels deep into a representation and uncover 15.6 times more embedded resources (70.9\% of which are available via onclick events).

\paragraph{Problem Description}
\label{problem}

The current rate of browsers implementing (and content authors adopting) client-side technologies such as JavaScript is much faster than crawlers' abilities to develop tools to crawl web resources that leverage the technologies. This leads to a difference between the web that crawlers can discover and the web that human users experience -- a challenge impacting the archives as well as other web-scale crawlers (e.g., those used by search engines). Over time, live web resources have been more heavily leveraging JavaScript (i.e., Ajax) to load embedded resources \cite{ijdl}. Because JavaScript-dependent representations are not accessible to web-scale archival crawlers, their representations are not fully archived. When the representation is replayed from the archive, the JavaScript will execute and may issue Ajax requests for a resource that is  on the live web, which leads to one of two possible outcomes: the live web ``leaking'' into the archive leading to an incorrect representation \cite{zombies}, or missing embedded resources (i.e., returns a 400 or 500 class HTTP response) in the archived resource leading to an incomplete representation, both of which result in reduced archival quality \cite{brunelleDamage}. When an archived deferred representation loads embedded resources from the live web via leakage, it is a \emph{zombie} resource, leaving the representation incorrect, and potentially \emph{prima facie violative} \cite{ainsworthFramework}.
We refer to the ease of archiving a Web resource as \emph{archivability}, and have shown that resources that rely on JavaScript to construct their representations have lower archivability than resources that avoid JavaScript \cite{ijdl}.

Heritrix archives pages by beginning with an initial seed list of Universal Resource Identifiers (URIs), dereferencing a URI from the list, archiving the returned representation, extracting embedded URIs to add to its crawl frontier, and repeating until the crawl frontier is exhausted. Heritrix does not execute any client-side scripts or use headless or headful browsing technologies\footnote{Even though it does not execute JavaScript, Heritrix v. 3.1.4 does peek into the embedded JavaScript code to extract links where possible \cite{htrixJS}.} \cite{googleJS}. 

We define \emph{deferred representations} as representations of resources that rely on JavaScript and other client-side technologies to initiate requests for embedded resources after the initial page load. We use the term \emph{deferred} because the representation is not fully realized and constructed until \emph{after} the JavaScript code is executed on the client. 
Note that the mere presence of JavaScript does not indicate that a representation will be deferred% -- a representation may have JavaScript that does not initiate Ajax requests for external resources
. A deferred representation may be interactive, but its reliance on Ajax and JavaScript to initiate HTTP requests for post-load resources makes the representation deferred.

HTTP transactions are stateless, meaning the representation is often indexable and identified by a combination of a timestamp and URI \cite{Fielding:2002:PDM:514183.514185}. However, 
client-side technologies such as JavaScript have made representations state\emph{ful}, allowing representations and their states to change independent of the URI and the time at which the representation was received from the server. In a resource with a deferred representation and multiple descendants\footnote{Descendants are defined in Section \ref{descs}.}, user or client-side interactions generate requests for additional embedded resources.

%Heritrix, like other web-scale crawlers, is not able to effectively crawl deferred representations, which leads to incorrect (i.e., leakage) or missing embedded resources (e.g., the missing splash page on the Wikipedia SOPA protest). 

%Headless browsing clients such as PhantomJS \cite{pjs} dereference URIs as would a browser (e.g., Google Chrome, Mozilla Firefox), allowing JavaScript to execute on the client. Using JavaScript to build a representation on page load is not the only challenge facing crawlers and archival tools. Content authors are also using JavaScript to construct interactive web \emph{applications}. These interactions may result in changes to the document object model (DOM) or initiate new HTTP GET requests for resources not yet loaded into the client. 
%We refer to a representation constructed as a result of user interaction or other client-side event without a subsequent request for the resource's URI as a \emph{descendant} (i.e., a member of the client-side event tree below the root)\footnote{The Dincturk Hypercube model refers to these as Ajax states or client-side states.}. Client-side events may also trigger a request, likely via Ajax, for additional resources to be included in the representation, which leads to deferred representations.

\paragraph{Contributions}
\label{contributions}

In this paper, we define a representation constructed as a result of user interaction or other client-side event without a subsequent request for the resource's URI as a \emph{descendant}\footnote{The Dincturk Hypercube model refers to these as Ajax states or client-side states.} (i.e., a member of the client-side event tree below the root). Client-side events may also trigger a request for additional resources to be included in the representation, which leads to deferred representations.

We explore the number and characteristics of descendants as they pertain to web archiving and explore the cost-benefit trade-off of actively crawling and archiving descendants en route to a higher quality, more complete archive. Dincturk et al. \cite{dincturkAjax} constructed a model for crawling Rich Internet Applications (RIAs) by discovering all possible descendants and identifying the simplest possible state machine to represent the states. We explore the archival implementations of their Hypercube model by discovering client-side states (represented as a tree) and their embedded resources to understand the impact that deferred representations have on the archives. %We also use the Hypercube model as the basis for our proposed two-tiered crawling approach.

%As part of our proposed two-tier crawling approach, we present a model for representing descendants, and construct a tree beginning with $s_0$ and expanding to descendants below $s_0$ for a sample set of resources, identifying the embedded resources required to complete their descendants. We also identify which embedded resources are not discovered by the crawler due to its inability to crawl descendants and deferred representations.

We evaluate the performance impacts of exploring descendants (i.e., crawl time, depth, and breadth) against the improved coverage of the crawler (i.e., frontier size) along with the presence of embedded resources unique to descendants in the Internet Archive, using Heritrix as our case study of a web-scale crawling tool. We show that the vast majority (92\% in $s_1$ and 96\% in $s_2$)\footnote{The terms $s_1$ and $s_2$ are defined in Section 3.} of embedded resources loaded as a result of user interactions are not archived, and that there are two levels in the interaction trees of our URI-Rs.

Throughout this paper we use Memento Framework terminology. Memento \cite{nelson:memento:tr, mementorfc} is a framework that standardizes Web archive access and terminology. Original (or live web) resources are identified by URI-R, and archived versions of URI-Rs are called \emph{mementos} and are identified by URI-M. 
\\\\

\section{Related Work}
\label{priorwork}
%Archivability helps us understand what makes representations easier or harder to archive; it can also provide an indication of how easy a page is to crawl. 
Banos et al. \cite{ipresArchivability} created an algorithm to evaluate archival success based on adherence to standards for the purpose of assigning an archivability score to a URI-R. In our previous work \cite{kellyTPDL2013}, we studied the impact of accessibility standards on archivability and memento completeness. We also measured the correlation between the adoption of JavaScript and the number of missing embedded resources in the archives \cite{ijdl}. 

Spaniol \cite{spaniol9catch, spaniol2009data, Denev:2009:SFQ:1687627.1687694} measured the quality of Web archives based on matching crawler strategies with resource change rates. Ben Saad and Gan\c{c}arski \cite{saad2011} performed a similar study regarding the importance of changes on a page. Gray and Martin \cite{mementoQuality} created a framework for high quality mementos and assessed their quality by measuring the missing embedded resources. In previous work \cite{brunelleDamage}, we assigned a quantitative metric to a previously qualitative measurement of memento quality and measured a reduction in memento quality caused by JavaScript. These works study quality, helping us understand what is missing from mementos.

Google has made efforts toward indexing deferred representations \cite{googleJS} -- a step in the direction of solving the archival challenges posed by JavaScript. Google's indexing focuses on rendering an accurate representation for indexing and discovering new URIs, but does not fully solve archival the challenges caused by JavaScript. Archiving web resources and indexing representations are different activities that have differing goals and processes.

Browsertrix \cite{browsertrix} and WebRecorder.io \cite{webrecorder} are page-at-a-time archival tools for deferred representations and descendants, but they require human interaction and are not suitable for web-scale archiving. Archive.is \cite{archivetoday} handles deferred representations well, but is a page-at-a-time archival tool and strips out embedded JavaScript from the memento. Stripping the embedded JavaScript leads to potentially reduced functionality in the memento and an inability to perform a post-mortem analysis of a page's intended behavior using the memento.

We proposed a two-tiered crawling approach for archiving deferred representations at web-scale that uses Heritrix and PhantomJS \cite{crawlingDeferred}. We measured the performance impact of incorporating a headless browsing utility in an archival workflow. Our work demonstrates that PhantomJS \cite{pjs} and its headless browsing approach can be used in conjunction with Heritrix to grow Heritrix's crawl frontier by 1.75 times and better archive deferred representations, but crawls 12.15 times slower than Heritrix alone. We build on this effort by enhancing the PhantomJS branch of the archival workflow to learn and execute interactions on the client with the Hypercube model. Note that PhantomJS cannot be used for all crawl targets because of the unacceptably slow crawl speed as compared to Heritrix. We use a classifier to identify which representations are deferred and require PhantomJS for complete archiving. 

Several efforts have studied client-side state. Mesbah et al. performed several experiments regarding crawling and indexing representations of web pages that rely on JavaScript \cite{mesbahCrawling, mesbahInferState} focusing mainly on search engine indexing and automatic testing \cite{mesbahTesting, mesbah2}. Singer et al. developed a method for predicting how users interact with pages to navigate within and between web resources \cite{hyptrails}. Rosenthal spoke about the difficulty of archiving representations reliant on JavaScript \cite{iipc2013, futureWeb}. Rosenthal et al. extended their LOCKSS work to include a mechanism for handling Ajax \cite{dshrDlib}. Using CRAWLJAX and Selenium to click on DOM elements with onclick events attached and monitor the HTTP traffic, they capture the Ajax-specific resources. While Rosenthal et al. measure performance based on the audits and repairs required, we focus on wall-clock time and frontier sizes to measure performance. Further, we omit form-filling, a feature that the LOCKSS enhancements provide. We extend this work to all interactions and study the depth of the interaction trees and best policies for crawling deferred representations.

These prior works investigate the archiving and crawling challenges that client-side technologies like JavaScript have introduced. In this work, we build on these past investigations to understand the multiple states that can be discovered on the client by mapping interactions and the additional resources required to build the descendants.
\\
\section{Descendant Model}
\label{descs}
Dincturk et al. \cite{dincturkAjax} present a model for crawling RIAs by constructing a graph of descendants\footnote{Dincturk et al. refer to these as ``AJAX states'' within the Hypercube model; we use a tree structure and therefore refer to these as descendants.}. A RIA is a resource with descendants and potentially a deferred representation. The work by Dincturk et al. focuses on Ajax requests for additional resources initiated by client-side events which leads to deferred representations with descendants. Their work, which serves as the mathematical foundation for our work, identifies the challenges with crawling Ajax-based representations and uses a \emph{hypercube strategy} to efficiently identify and navigate between all client-side states of a deferred representation. Their model defines a client-side state as a state reachable from a URL through client-side events and is uniquely identified by the state's DOM. That is, two states are identified as equivalent if their DOM (e.g., HTML) is directly equivalent.

The hypercube model is defined by the finite state machine (FSM) 
\\\\
\noindent$M = (S, s_0, \Sigma, \delta)$ 
\\\\
\noindent and defined further in Equation \ref{statedef}, where

\begin{itemize}
\item $S$ is the finite set of client states
\item $s_0 \in S$ is the initial state reached by dereferencing the URI-R and executing the initial on-load events
%\item $\Sigma$ is the set of all events in the application 
\item $e \in \Sigma$ defines the client-side event $e$ as a member of the set of all events $\Sigma$
\item $\delta : S \text{x} \Sigma \rightarrow S$ is the transition function in which a client-side event is executed and leads to a new state
\end{itemize}
%- $S$ is the \emph{finite set of client states}\\
%- $s_1 \in S$ is the \emph{initial client state of the URL}\\
%- $\Sigma$ is the set of all events in the application \\
%- $e \in \Sigma$ defines the client-side event $e$ as a member of the set of all events $\Sigma$\\
%- $\delta : S \text{x} \Sigma \rightarrow S$ is the \emph{transition function} in which a client-side event is executed and leads to a new state (or, in our model, a sub-representation)

\begin{equation}
\begin{split} 
&s_i, s_j \in S\\
&\delta(s_i, e) = s_j\\
&e = \text{client-side event}\\
&j = i+1
\end{split}
\label{statedef}
\end{equation}

Dincturk et al. define a graph $G=(V, E)$ in which $V$ is the set of vertices $v_i \in V$ where $v_i$ represents an ``AJAX State'' $s_i$. Edges represent the transitions, or events, $e$ such that $(v_i, v_j;e) \in E$ IFF $\delta(s_i, e)=s_j$. A path $P$ is a series of adjacent edges that constitute a series of transitions from $s_0$ to $s_i$ via $e_{i...j}$. In effect, $P$ is a series of descendants derived from $s_0$ with one descendant at each level of the tree.

We adopt the FSM presented by Dincturk et al. nearly in its entirety. Because our application of this FSM is web archiving, our goal is to identify all of the embedded resources required by the representation to build any descendant as a result of user interactions or client-side events, archive them, and be able to replay them when a user interacts with the memento. 

The representation returned by simply dereferencing a URI-R is defined as $URI$-$R_{s_0}$. Subsequent descendants $URI$-$R_{s_i}$ and $URI$-$R_{s_j}$ are derived from $URI$-$R_{s_0}$ through a series of events $e_{i...j} \in \Sigma$. We define a descendant $URI$-$R_{s_i}$ as a client-side state originating at $URI$-$R_{s_0}$ as transitioning via events $e$ such that $\delta(s_0, e) = s_1$. Additionally, we define our paths through $G$ as the set of embedded resources required to move from $s_0$ to $s_i$.

\begin{figure*}[hT]
\centering
    \includegraphics[width=0.65\textwidth]{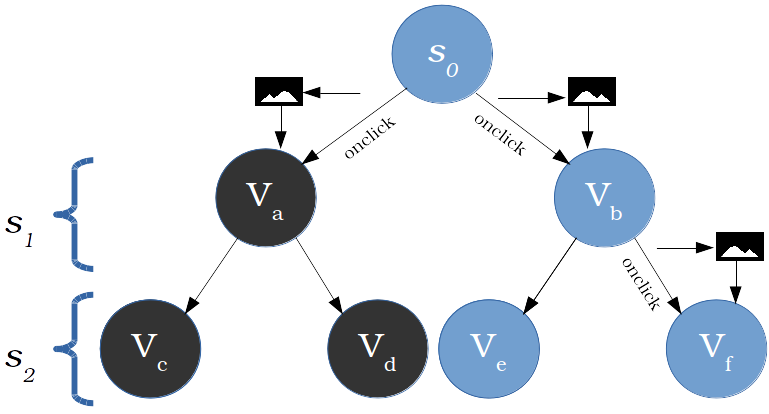}
  \caption{A generic, three-level client-side state tree with interactions as state transitions.}
\label{diagram}
\end{figure*}

We present a generic interaction tree of descendants in Figure \ref{diagram}. When we dereference a URI-R, we get a representation from the server; this is $s_0$. If there are two interactions available from $s_0$, we can execute the interactions to get to $V_a$ or $V_b$ from our root $s_0$ (note that the onclick event required an external image to be retrieved). In this example, $V_a$ and $V_b$ are descendants of $s_0$ and are both $s_1$ in $P$ from $s_0$. If new interactions are available from $V_a$, we can reach $V_c$ and $V_d$, which are both $s_2$ in $P$ from $s_0$ (similarly, we can reach $V_e$ and $V_f$ from $V_b$, peers of $V_c$ and $V_d$).

Because of the differences between our model and the hypercube model (Section \ref{paths}), our focus on web archiving, and to ensure we have omniscient knowledge of all interactions, state dependencies, equivalences, and available interactions, we organize the states as a tree rather than a hypercube. Because new interactions lead to states $s_n$ deeper in the tree, we generalize the levels of the trees as $s_n$ and refer to new states by their vertices $V_n$.

% A category with the (minimum) three required fields

\section{State Equivalency}
\label{paths}
Due to the archival focus of this study, we have a different concept of state equivalence than the Hypercube model. While Dincturk establishes state equivalence based on the DOM (using strict equivalence based on string comparison), we consider the embedded resources required to construct a descendant. We consider states to be equivalent if they rely on the same embedded resources. As such, we define the set of embedded resources for a descendant $s_n$ as $R_n$. 

Any two states with identical unordered sets of embedded resources are defined as equivalent, effectively a bijection between the two states. $P$ between $s_0$ to $s_i$ is isomorphic if, over the course of $P$, the embedded resources required are identical, and each state within $P$ are bijections. Note that in Figure \ref{diagram}, $V_a$, $V_c$, and $V_d$ are equivalent because they require the same set of embedded resources, even if $V_c$ and $V_d$ are reached through an additional $e_i$ and $e_j$ from $V_a$.

Two paths are identical if, over the course of each $s_n$ $\in$ $P$, the cumulative set of embedded resources required to render each descendant is identical. We define the set of embedded resources over the entire path as $RP$ in Equation \ref{resourcepaths}.

\begin{equation}
RP = \sum_{i=0}^{n \in P} R_n
\label{resourcepaths}
\end{equation}

We present our process for traversing paths in Algorithm \ref{algo1}. We traverse all states within the interaction tree to understand what embedded resources are required by each state. If a state $s$ requires a new embedded resource that has not yet been added to the crawl frontier, it is added as part of path $P$. From $RP$, we identify the archival coverage (using Memento -- line 10). We also identify the duplicate URI-Rs by canonicalizing, trimming fragment identifiers from the URI-R, and using string comparisons to determine equality.

\begin{algorithm}
\ForAll{ URI-R}{
 find $s_0$...$s_n$\;
 construct tree G of all progressions\;
 traverse G; identify $R_n$ of $s_n$\;
 \If{ $s_n$ has a new resource}{
  treat $s_n$ as part of $P$\;}
}
identify $RP$\;
find URI-Ms of $RP$ to determine coverage\;
de-duplicate $RP$ to determine overlap\;

 \caption{Algorithm for traversing $P$.}
  \label{algo1}
\end{algorithm}
%%%%$

\begin{figure*}[hp]
\centering
\includegraphics[width=0.98\textwidth]{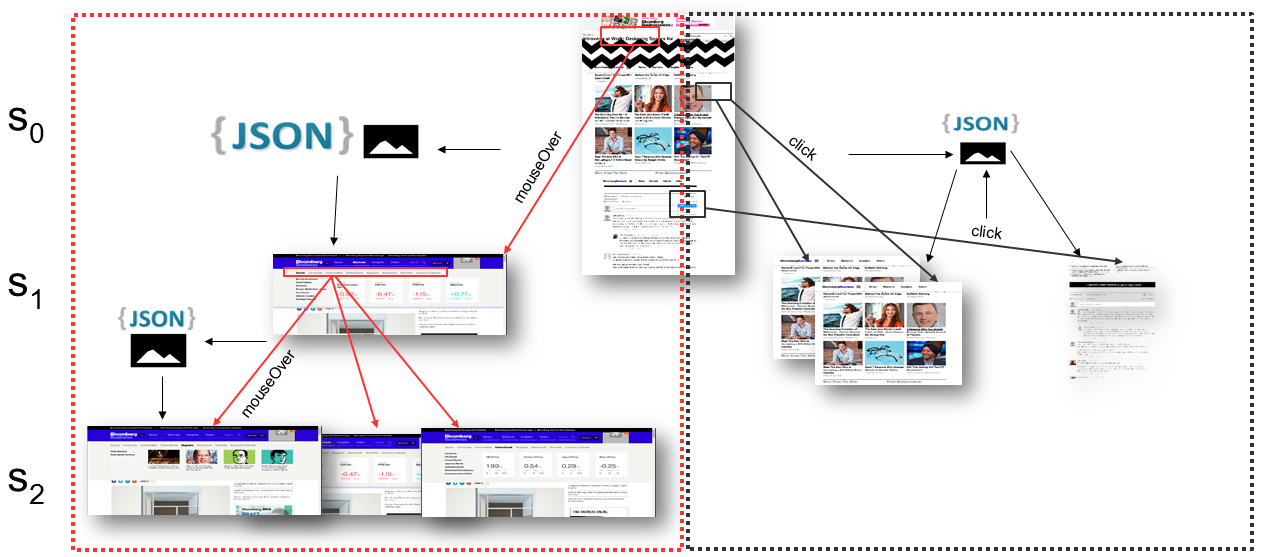}
\caption{Example state tree of \texttt{http://www.bloomberg.com/bw/articles/2014-06-16/open -plan-offices-for-people-who-hate-open-plan-offices}. Mouseover events lead to multiple descendants at $s_1$ and further mouseover events lead to descendants at $s_2$, each requiring Ajax requests for JSON and image resources. Figures \ref{realStateL} and \ref{realStateR} provide closer views of this figure.}
\label{realState}
\end{figure*}

\begin{figure*}[hp]
\centering
\includegraphics[width=0.98\textwidth]{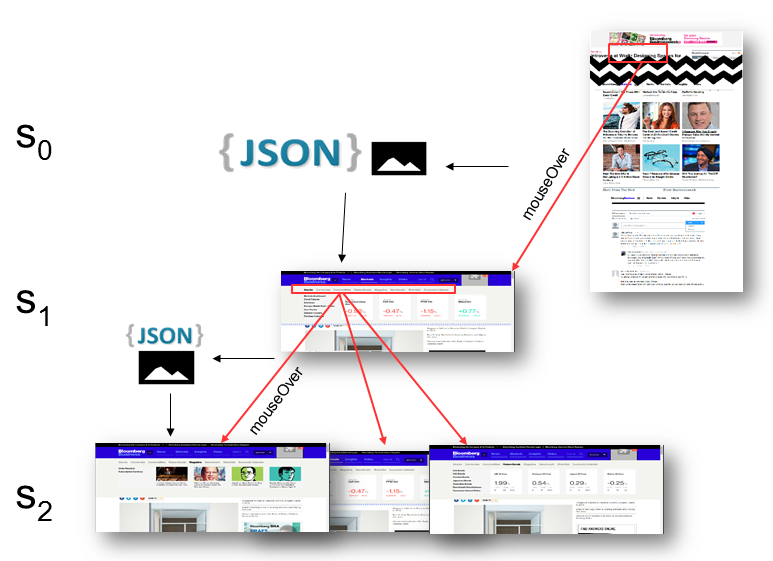}
\caption{This figure highlights the menu and submenu descendants from the left side of Figure \ref{realState}.}
\label{realStateL}
\end{figure*}

As an example, we present the state tree of the top level page at Bloomberg.com in Figure \ref{realState}. At $s_0$, the page has a menu at the top of the page with a mouseover event listener (Figure \ref{realStateL}). Mousing over the labels initiates Ajax requests for JSON data, and the data is used to populate a sub-menu ($s_1$). The sub-menu has another mouseover menu that requests images and other JSON data to display new information, such as stock market data and movie reviews ($s_2$). Note that $s_1$ and $s_2$ are very broad given the number of menu items. This is an example of $P$ through two levels of mouseover interactions that leads to new JSON and image embedded resources. 

\begin{figure*}[hp]
\centering
\includegraphics[width=0.98\textwidth]{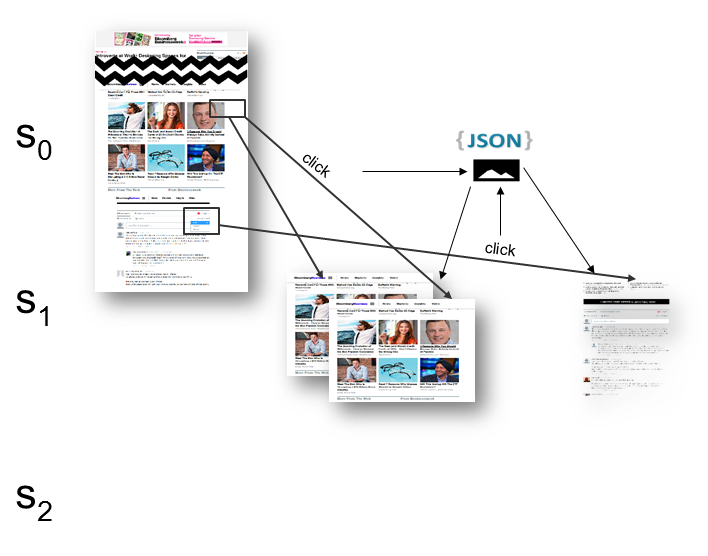}
\caption{This figure highlights the descendants created though click events from the right side of Figure \ref{realState}.}
\label{realStateR}
\end{figure*}

The page also has onclick events (Figure \ref{realStateR}). These onclick events also lead to descendants at $s_1$, but not $s_2$. However, the onclick events lead to equivalent descendants that we identify as equivalent.

Finally, note that the comments section has a form that users can fill out to enter a comment. This is beyond the scope of this work; this type of action changes the representation itself and can lead to infinite states, changing the nature of our investigation from identifying archival targets to altering the live web record.

\section{Approach}
\label{approach}

To measure descendants, we needed to construct a tool that can crawl, uncover, and understand descendants and deferred representations. We have previously shown that PhantomJS is an effective utility for crawling deferred representations \cite{crawlingDeferred}. 
We constructed a PhantomJS-based utility that dereferences a URI-R, identifies the interactive portions of the DOM (i.e., the DOM elements with event listeners), and constructs a tree of descendants, reached via initiating interactions and client-side events (just as in the Hypercube model). PhantomJS records the set of embedded resources requested by the client; in a production system, this would be the set of resources added to the Heritrix crawl frontier.

Because PhantomJS is closely tied to the DOM and client's JavaScript engine, race conditions and other event listener assignments prevents PhantomJS from understanding the entirety of events available on a representation. As such, we leveraged VisualEvent\footnote{\url{https://github.com/DataTables/VisualEvent}}, a bookmarklet that is designed to visually display the DOM elements that have event listeners and JavaScript functions attached \cite{seleniumpjs}, to understand which events and interactions can be executed on the client \cite{seleniumpjs}. 
Our PhantomJS tool uses the list of events identified by VisualEvent to construct a set of interactions $E$ that may lead to descendants. PhantomJS performs an exhaustive depth-first traversal of all possible combinations of interactions. Post-mortem, we perform state equivalence and identify the number of unique paths $P$, states $s_n$, and embedded resources $RP$ that a crawler would have to visit in order to comprehensively archive the resources needed to provide full functionality in a memento.

We use the same 440 URI-R dataset\footnote{\url{https://github.com/jbrunelle/DataSets/blob/master/440uris.txt}} from our prior investigation of crawling deferred representations \cite{crawlingDeferred}. We generated URI-Rs by randomly generating Bitly strings and identifying their redirection targets. We used PhantomJS to identify each URI-R as having a deferred or nondeferred representation and identify the number and type of descendants and interactions available on the representations of URI-Rs in this set, along with the descendants within the interaction tree and embedded resources required to build the descendant representations. 

Note that a $V_n$ is reached by a series of interactions $e_{i...j}$. We consider a descendant that is a candidate to add to the tree identical to another descendant within the tree if the set of interactions to reach the descendant are identical. If we encounter a potential descendant that is reachable by the same interactions as another descendant within the tree, we do not add the descendant to the tree because the descendant already exists within the tree\footnote{Note that this refers to equivalency of interaction scripts, meaning the crawler should not visit this state, rather than two states that are reached with different interactions but have the same sets of embedded resources (Section \ref{paths}).}.

We present our algorithm for crawling the descendants in Algorithm \ref{algo2}. We begin by using PhantomJS to dereference a URI-R (line 4) at $s_0$, and use VisualEvent to extract the interactive elements (line 5). We identify all possible combinations of interactions and use them as an interaction frontier (line 7), and iterate through the interaction frontier to crawl $s_1$. From $s_1$, we extract all possible interactions available and add them to the interaction frontier. We iterate through the interaction frontier until we have exhausted all possible combinations of interactions at each $s_n$. At the end of each $s_n$ construction, we run state deduplication (line 14). We deem two interaction scripts as equivalent if they perform identical actions in identical order:
\\ \\
\noindent$\{e_{i}, e_{i+1}, ..., e{i+n}\}$ = $\{e_{j}, e_{j+1}, ..., e{j+n}\}$

\begin{algorithm}
\ForAll{ URI-R}{
 run state $s_0$\;
 \Begin{
  run PhantomJS; monitor and log $R_n$\;
  call VisualEvent functions to retrieve
    interactive DOM elements and $e_{i...j}$\;
    }
 construct interaction scripts
 from all possible combinations of available 
    interactions to read $s_n$\;
 \ForAll {interaction scripts}{
  run PhantomJS, executing interactions 
    in script\;
  monitor and log $R_n$\;
  call VisualEvent for events to 
  	reach $s_{n+1}$\;
  construct new $s_{n+1}$\;
  push new interaction scripts to list of
     all interaction scripts\;
  run interaction script de-duplication
  \Begin{
   \If {$s_{n+1}$ has identical 
     series of interactions in G}{
     remove $s_{n+1}$
     }
     }
     }
}
 \caption{Algorithm for constructing $G$.}
  \label{algo2}
\end{algorithm}

\section{Edge Cases}
\label{edge}
The approach that we identify in Section \ref{approach} is suitable for most of the deferred representations that a web user may encounter while browsing. However, deferred representations with certain conditions are not handled by our approach. Some representations use a DIV overlayed on the entire window area and identify interactions and events according to the pixel the user clicks. This creates an interaction frontier of 
$(\text{Width} \times \text{Height})!$ or 2,073,600! for a screen size of $1920 \times 1080$ pixels. Due to this massive frontier size, we omit such interactions. Mapping (e.g., Google Maps)  and similar applications that might have a near-infinite descendants are outside the scope of this  work.

For these style of deferred representations, a \emph{canned} set of interactions (e.g., pan once, zoom twice, right click, pan again) would be more useful \cite{seleniumpjs}. With enough of these canned interactions, a sizable portion of the descendants can be identified by a crawler over time, with coverage scaling with the number of executions performed. This is the archival equivalent of the Halting Problem -- it is difficult to recognize when the crawler has captured \emph{enough} of the embedded resources, when it should stop, or when it has captured everything.

\section{Descendant States}
\label{states}
During our crawl of the 440 URI-Rs, we classified each as having a deferred or nondeferred representation. As previously discussed, URI-Rs with deferred representations will have an event that causes the JavaScript and Ajax in the representation to request additional resources. We dereferenced each of our 440 URI-Rs and identified 137 URI-Rs with nondeferred representations and 303 URI-Rs with deferred representations. 

\subsection{Dataset Differences}
\label{diffs}

\begin{table}
\centering
\begin{tabular}{ l | r | r | r | r }
\multirow{2}{*}{\textbf{Event Type}} & \multicolumn{2}{c}{\textbf{Deferred}} & \multicolumn{2}{c}{\textbf{Nondeferred}} \\
                            & Average         & $s$         & Average   &     $s$    \\
\hline
\hline
Depth & 0.47  & 0.5 & 0 & 0 \\
\hline
Breadth & 36.16  & 97.15 & 0.53 & 2.62 \\
\hline
Descendants & 38.5  & 780.72 & 0.62 & 2.81 \\
\hline
\end{tabular}
  \caption{The average distribution of descendants within the deferred representation URI-R set.}
  \label{depthvbreadth}
\end{table}

The nondeferred URI-Rs have a much smaller graph of descendants, and therefore we expect them to be easier to crawl. The nondeferred representation set of URI-Rs had $\overline{|S_{descendants}|}=0.62$ per URI-R ($s=2.81$, $M=101$)\footnote{We refer to the standard deviation and median values of our observed sample set as $s$ and $M$, respectively.} as shown in Table \ref{depthvbreadth}. Nondeferred representations have a depth (i.e., max length of the $P$) of 0 (after state deduplication) since there are no new states reached as a result of the first round of interactions (that is, the set of interactions available in the initial representation does not grow as a result of subsequent interactions). Nondeferred representations have descendants at $s_1$ but the descendants do not result in additional embedded resources. However, there are 0.53 interactions or events in $s_0$ that, without our \emph{a priori} knowledge of the dataset, may lead to new states or event-triggered requests for new embedded resources.

\begin{table}
\centering
\begin{tabular}{ l | r | r }
\textbf{\# States} & \textbf{Deferred} & \textbf{Nondeferred} \\
\hline
\hline
Min & 0 & 0\\
\hline
Max & 7,308 & 13\\
\hline
Median & 1 (occurrences 17) & 0 (occurrences 119)\\
\hline
\end{tabular}
  \caption{The range of descendants varies greatly among the deferred representations.}
  \label{stateminmedmod}
\end{table}

Deferred representations are much more complex, with $\overline{|S_{descendants}|}=38.5$ per URI-R ($s=780.72$). The standard deviation of the sample is quite large, with the number of states varying greatly from resource to resource. For example, the maximum number of descendants for a URI-R is 7,308 (Table \ref{stateminmedmod}). Further, there are many interactions available in deferred representations (36.16 per URI-R). Surprisingly, deferred representations are relatively \emph{shallow}, with an average depth of 0.47 levels beyond the first set of interactions per URI-R ($s=0.5$) and a maximum path depth of 2. This is counter to our initial intuition that deferred representations would have large, deep trees of interactions to traverse to retrieve all of the possible embedded resources for all descendants\footnote{Our dataset and toolset omits edge cases as described in Section \ref{edge}.}.

%%%%the below table was generated with:
%%perl reclassifyDeferred.pl deferred_uris.txt nondeferred_uris.txt def3out3.txt nondef3out3.txt deferred_run3/ nondeferred_run3/
%%perl collectStateStats2.pl nondef3out3.txt nondefrun3.txt
%%perl collectStateStats2.pl def3out3.txt defrun3.txt

\begin{table}
\centering
\begin{tabular}{p{2cm} | r | r | >{\raggedleft\arraybackslash}p{2cm}}
\multirow{2}{*}{\textbf{Event Type}} & \multicolumn{2}{c}{\textbf{Percent of URI-Rs}} & \textbf{Contribution}\\
                            & Deferred         & Nondeferred       &    \textbf{to $RP_{new}$}    \\
\hline
\hline
click & 62.11\% & 4.29\% & 63.2\%\\
\hline
mouseover & 25.26\% & 3.00\% & 4.7\%\\
\hline
mousedown & 16.84\% & 1.72\% & 2.8\%\\
\hline
blur & 14.74\% & 0.86\% & 9.8\%\\
\hline
change & 11.58\% & 2.14\% & 0.0\%\\
\hline
mouseout & 8.42\% & 0.00\% & 0.8\%\\
\hline
submit & 6.32\% & 0.00\% & 0.0\%\\
\hline
unload & 5.26\% & 0.00\% & 1.2\%\\
\hline
keydown & 4.21\% & 0.00\% & 0.2\%\\
\hline
focus & 4.21\% & 0.00\% & 0.0\%\\
\hline
keypress & 2.11\% & 0.00\% & 5.5\%\\
\hline
focusout & 1.05\% & 0.00\% & 0.0\%\\
\hline
dblclick & 1.05\% & 0.00\% & 0.0\%\\
\hline
submit & 0.10\% & 0.43\% & 0.9\%\\
\hline
mouseup & 0.00\% & 0.86\% & 0.0\%\\
\hline
focus & 0.00\% & 0.43\% & 0.0\%\\
\hline
other & 29.47\% & 0.86\% & 11.0\%\\
\hline
\end{tabular}
  \caption{Breakdown of the URI-Rs with various events attached to their DOMs and the percent of all new embedded resources contributed by the events.}
  \label{events}
\end{table}

%\begin{figure}
%\centering
%\includegraphics[width=0.48\textwidth]{./newByEventHorz.png}
%\caption{Number of new embedded resources requested as a result of various events.}
%\label{newbyevent}
%\end{figure}

The types of events on the client also vary depending on the event that is executed to create the new $s_n$. For example, onclick events are prevalent in URI-Rs with deferred representations, with 62.11\% of all URI-Rs containing an onclick event (Table \ref{events}). Even in the nondeferred set of URI-Rs, 4.29\% of the URI-Rs have an onclick event attached to their DOM. While other events occur with relative frequency, clicks dominate the initiated requests for additional embedded resources in deferred representations (Table \ref{events}), with onclick events being responsible for initiating the requests for 63.2\% of new embedded resources (and, by definition, 0\% in the nondeferred representation set). Recall that Rosenthal et al. are using only click interactions to interact with pages. Table \ref{events} suggests that their approach is effective considering most events are onclick events and the highest value target event (i.e., the most embedded resources are discovered through onclick events).

\subsection{Traversing Paths}
\label{traversingPaths}
As we discussed in Section \ref{paths}, $P$ identifies a unique navigation through descendants to uncover the URI-Rs of new embedded resources. In our dataset, we uncovered 8,691 descendants (8,519 for the deferred set, 172 for the nondeferred set) as a result of client-side events, which is 19.7 descendants per URI-R. However, we only identified 2,080 paths through these descendants to uncover all of the new embedded resources, which is 4.7 paths per URI-R.

\begin{figure}
\centering
\includegraphics[width=0.48\textwidth]{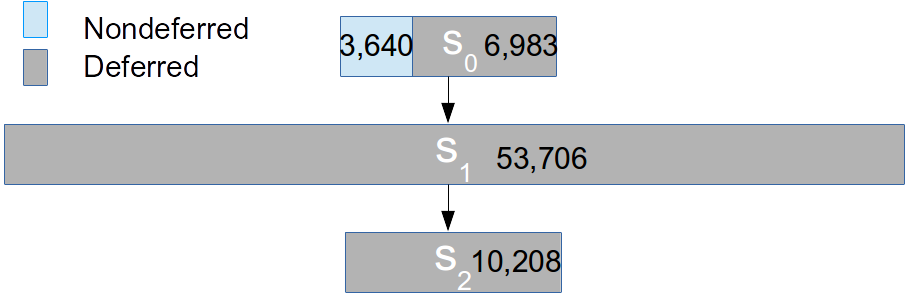}
\caption{Crawling $s_1$ provides the greatest contribution to $RP$; the additions to the crawl frontier by $s_0$ (10,623) and $s_2$ (10,208) only differ by 415 URIs.}
\label{growth2a}
\end{figure}

Nondeferred representations have more embedded resources ($R_0$= 31.02 per URI-R) than their deferred counterparts ($R_0$= 25.39 per URI-R) at their initial $s_0$. The paths $P$ through the descendants are responsible for uncovering 54,378 new embedded resources (out of 66,320 total). That is, $|R_0|$=11,942 (7,692 from the deferred representations and 4,350 from the nondeferred representations), $|R_0 + R_1|$=56,957, and $|R_0 + R_1 + R_2|$=$|RP|$=66,320. This shows that traversing $P_n$ to reach $s_1$ and $s_2$ will significantly increase the crawl frontier beyond the base case of $s_0$, but crawling $s_1$ provides larger contributions to the frontier than both $s_0$ and $s_2$. 
As we mentioned in Section \ref{diffs}, the depth of the deferred representations was surprisingly shallow (max($|P|$)=2). However, the overwhelming majority of the URI-Rs added to the crawl frontier were identified by exploring the paths $P$ of the descendants (Figure \ref{growth2a}). %Note that the total crawl frontier at each depth remains constant at 4,250 URI-Rs for nondeferred representations because they do not require additional resources as a result of traversing deeper into their paths.

%\begin{figure}
%\centering
%\includegraphics[width=0.48\textwidth]{./growth1.png}
%\caption{The crawl frontier grows as $P$ is followed except in the case of nondeferred representations. However, the number of new URI-Rs discovered at depth two is much less than that at depth one.}
%\label{growth1}
%\end{figure}

Note that out of the total 8,691 total descendants, the nondeferred representations have only 138 occurrences of $s_0$ and 34 occurrences of $s_1$. The deferred representations have 6,051 occurrences of $s_1$ and 2,468 occurrences of $s_2$. Following $P$ through each $s_1$ adds $R_1$=53,706 URI-Rs to the crawl frontier, or 8.88 URI-Rs per descendant. This shows that deferred representations have many more descendants than nondeferred representations. Since $R_2$=10,208, we add, on average, 4.14 new URI-Rs to the frontier per $s_2$ followed in $P$. According to these averages, crawling $s_1$ provides the largest benefit to the crawl frontier. If we consider only the 2,080 paths that lead to new embedded resources, we would add 30.73 URI-Rs to the crawl frontier per $P$.

\subsection{Impact on Crawl Time}
\label{value}

Our prior work \cite{crawlingDeferred} measured crawl times for Heritrix and PhantomJS, including deferred and nondeferred representations. %We measured that Heritrix crawls 12.56 URIs/second, while PhantomJS crawls 0.255 URIs/second. 
We measured that Heritrix is 12.15 times faster than PhantomJS (2.065 URIs/second versus 0.170 URIs/second, respectively). 
Using these results and the set of states $S$ that PhantomJS can uncover and visit, we calculate the expected frontier size and crawl time that we can expect during a crawl of our 440 URI-Rs. 
Using these metrics, we calculated the $s_0$ crawl time for Heritrix-only crawls, PhantomJS crawls of only the URI-Rs with deferred representations, and $s_1$ and $s_2$ uncovered by our PhantomJS utility. %We plot this against our frontier size in the crawl in Figure \ref{roi}.

%\begin{figure}
%\centering
%\includegraphics[width=0.48\textwidth]{./ROIint.png}
%\caption{Crawl time versus added frontier size.}
%\label{roi}
%\end{figure}

As we note in Table \ref{relatives}, $s_1$ has the greatest addition to the crawl frontier. If we omit $s_2$ from our crawl, we still discover 82\% of the embedded resources required by our deferred representations and reduce crawl time by 30\%. Depending on the goal of the crawl, different crawl policies would be optimal when crawling deferred representations. A policy to optimize archival coverage and memento quality should traverse each $s \in S$ and maximize $RP$; we will refer to this policy as the \emph{Max Coverage} crawl policy. Alternatively, a crawl policy with the goal of optimizing return-on-investment (ROI) should optimize the crawl time versus frontier size; we refer to this policy as the \emph{Max ROI} crawl policy. For maximizing ROI, we recommend a crawl policy that omits $s_2$ and instead uses Heritrix and PhantomJS to crawl $s_0$ and $s_1$, since $s_1$ provides the greatest contribution to $RP$ per crawl time (Table \ref{relatives}).

As shown in Table \ref{relatives}, using the \emph{Max Coverage} policy will lead to a crawl time 38.9 times longer than using only Heritrix to perform the crawl, but will also discover and add to the crawl frontier 15.60 times more URI-Rs. Alternatively, the \emph{Max ROI} policy will have a crawl time 27.04 times longer than Heritrix-only crawls, but will add 13.40 times more URI-Rs to the frontier.

\begin{table}
\centering
\begin{tabular}{p{2.5cm} | p{0.75cm} | r | r | r}
 & \textbf{H-only} & \textbf{$s_0$} & \textbf{$s_1$} & \textbf{$s_2$}\\
\hline
\hline
\textbf{Time (s)} & 1,035 & 8,452 & 27,990 & 40,258\\
\hline
\textbf{Size (URI-Rs)} & 4,250 & 11,942 & 56,957 & 66,320\\
\hline
\hline
\textbf{Time Increase} & - & 8.12x & 27.04x & 38.90x\\
\hline
\textbf{Size Increase} & - & 2.81x & 13.40x & 15.60x\\
\hline
\textbf{New URIs per added second of crawl time} & - & 1.04 & 2.30 & 0.76\\
\hline
\end{tabular}
  \caption{The increases in run time and frontier size relative to the Heritrix-only (H-only) run. Note that we use the \emph{Max Coverage} strategy to generate these performance results.}
  \label{relatives}
\end{table}

\section{Archival Coverage}

While the increases in frontier size as presented in Section \ref{states} appear impressive, we can only identify the impact on the archives' holdings by identifying which embedded resources have mementos in today's archives. We used Memento to retrieve the TimeMap of each embedded resource's URI-R to determine whether the embedded resource had any mementos (i.e., has been archived before) or if the resource identified by the URI-R has not been previously archived.

The embedded resources from our entire set of 440 URI-Rs are very well archived -- only 12\% of the set of embedded resources in $s_0$ do not have a memento. This is consistent with the archival rates of resources from our prior studies \cite{hmotwia, ijdl}. 
We only consider the \emph{new} embedded resources in the descendants in $s_1$ and $s_2$. That is, we only consider the embedded resources added to the descendant that were not present in the previous state, or the set of resources $R_{n+1}$ not a subset of the previous state's $R_n$. More formally, we define new embedded resources $R_{\text{new}}$ in Equation \ref{newER}.

\begin{equation}
 R_{\text{new}} = \forall r \in (R_{n+1} - R_n), n \geq 0
\label{newER}
\end{equation}

\begin{figure}
\centering
\includegraphics[width=0.48\textwidth]{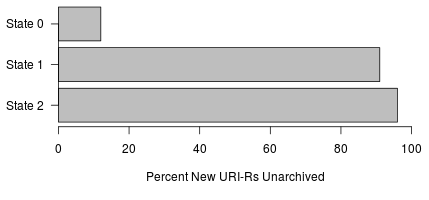}
\caption{Embedded resources discovered in $s_1$ and $s_2$ are much more frequently unarchived (92\% and 96\%, respectively) than $s_0$ (12\% unarchived).}
\label{depth}
\end{figure}

However, the archival coverage of $s_1$ and $s_2$ is much lower, with 92\% of $R_{\text{new}}$ in $s_1$ missing from the archives (i.e., the URI-R of the embedded resource does not have a memento), and 96\% of $R_{\text{new}}$ in $s_2$ missing from the archives. This demonstrates that the embedded resources required to construct descendants are not well archived. Because $s_0$ is highly visible to crawlers such as Heritrix and archival services like Archive.is \cite{archivetoday}, it is archived at a much higher rate than the descendants (Figure \ref{depth}).

\begin{figure}
  \begin{center}
    \subfigure[Images are most frequently unarchived in deferred representations.]{\label{mimes}\includegraphics[width=0.45\textwidth]{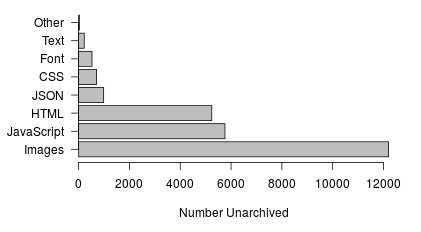}}\\
    %\subfigure[Unarchived image sizes are, on average, 3.5KB.]{\label{mimesizes}\includegraphics[width=0.45\textwidth]{./mimeSizeHisto.png}}
    \subfigure[Unarchived images vary in size, with most being small and a few being very large.]{\label{mimesizes}\includegraphics[width=0.45\textwidth]{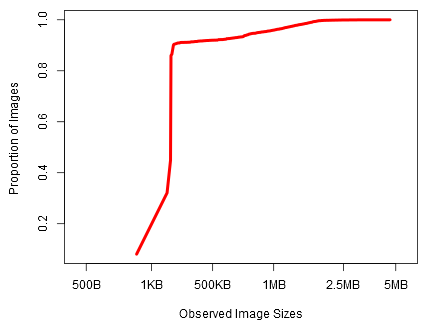}}
  \end{center}
  \caption{Images, JavaScript, and HTML are the most frequently occurring unarchived resources in deferred representations, with some unarchived images being quite large.}
  \label{mimeFigs}
\end{figure}

In deferred representations, the unarchived embedded resources are most frequently images (Figure \ref{mimes}), with additional JavaScript files edging out HTML as the second most frequently unarchived MIME-type. The unarchived images specific to deferred representations (shown in the Cumulative Distribution Function (CDF)\footnote{The CDFs in this paper illustrate the proportion of the observations (y-axis) that are equal to or less than the value on the x-axis. For example, Figure \ref{mimesizes} shows that most unarchived images are small (less than 500KB in size) as shown by the sharp increase in proportion of images (y axes) to 91\% before the 500KB tic on the x-axis.} in Figure \ref{mimesizes}) vary in size between near 0B to 4.6MB, and average 3.5KB. The majority of images are small in size but several are quite large and presumably important (according to our prior importance metric $D_m$ \cite{damageIJDL}).

%\begin{figure}
%\centering
%\includegraphics[width=0.5\textwidth]{./unarchivedByMime.png}
%\caption{Images are most frequently unarchived in deferred representations.}
%\label{mimes}
%\end{figure}

%\begin{figure}
%\centering
%\includegraphics[width=0.5\textwidth]{./mimeSizeHisto.png}
%\caption{Unarchived image sizes are, on average, 3.5KB.}
%\label{mimesizes}
%\end{figure}

\begin{table}
\centering
\begin{tabular}{ p{5cm} | r}
\textbf{URI-R} & \textbf{Occurrences} \\
\hline
\hline
\url{ads.pubmatic.com/AdServer/js/showad.js#PIX&kdntuid=1&p=52041&s=undefined&a=undefined&it=0} & 1782 \\
\hline
\url{edge.quantserve.com/quant.js} & 1656 \\
\hline
\url{www.benzinga.com/ajax-cache/market-overview/index-update} & 1629 \\
\hline
\url{ads.pubmatic.com/AdServer/js/showad.js} & 1503 \\
\hline
\url{www.google-analytics.com/analytics.js} & 1330 \\
\hline
\url{b.scorecardresearch.com/beacon.js} & 1291 \\
\hline
\url{www.google-analytics.com/ga.js} & 1208 \\
\hline
\url{www.google.com/pagead/drt/ui} & 1151 \\
\hline
\url{js.moatads.com/advancedigital402839074273/moatad.js} & 1112 \\
\hline
\url{a.postrelease.com/serve/load.js?async=true} &  907\\
\hline
\hline
Total &  12,239\\
\hline
\end{tabular}
  \caption{The top 10 URI-Rs that appear as embedded resources in descendants make up 22.4\% of all resources added to the crawl frontier. }
  \label{counts}
\end{table}

%\begin{figure}
%\centering
%\includegraphics[width=0.48\textwidth]{./growth2a.png}
%\caption{}
%\label{growth2a}
%\end{figure}

%\begin{figure}
%  \begin{center}
%    \subfigure[Each descendant contributes to the overall frontier size, with $s_1$ providing the greatest contribution and $s_0$ and $s_2$ providing approximately equal contributions.]{\label{growth2a1}\includegraphics[width=0.48\textwidth]{./growth2a_split1.png}}\\
%    \subfigure[Deferred representations contribute to most of the unarchived embedded resources.]{\label{growth2a2}\includegraphics[width=0.48\textwidth]{./growth2a_split2.png}}  
%  \end{center}
%  \caption{The number of new URI-Rs discovered at $s_2$ is much less than that at $s_1$. Further, the archival coverage reduces deeper into the tree.}
%  \label{growth2a}
%\end{figure}

%\begin{figure}
%  \begin{center}
%    \subfigure[Embedded resources discovered in $s_1$ and $s_2$ are much more frequently unarchived (92\% and 96\%, respectively) than $s_0$ (12\% unarchived).]{\label{growth2a2}\includegraphics[width=0.48\textwidth]{./archivedByDepthPctHorz.png}}\\
%    \subfigure[Crawling $s_1$ provides the greatest contribution to $RP$; the additions to the crawl frontier by $s_0$ (10,623) and $s_1$ (10,208) only differ by 415 URIs.]{\label{growth2a1}\includegraphics[width=0.48\textwidth]{./tree_diagram.png}}  
%  \end{center}
%  \caption{The number of new URI-Rs discovered at $s_2$ is much less than that at $s_1$. Further, the archival coverage reduces deeper into the tree.}
%  \label{growth2a}
%\end{figure}

\begin{figure}
\centering
\includegraphics[width=0.48\textwidth]{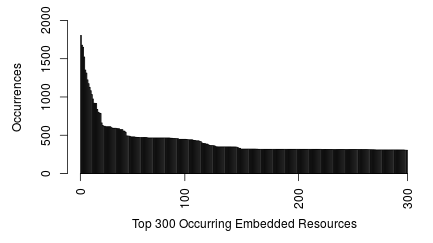}
\caption{The occurrence of embedded resources loaded into deferred representation descendants.}
\label{longtail}
\end{figure}

We also observe a large amount of overlap between the embedded resources among descendants. For example, the top 10 embedded resources and their occurrence counts are provided in Table \ref{counts} (we trim the session-specific portions of the URI-Rs for comparison purposes). In all, just the top 10 occurring embedded resources account for 22.4\% of $R_{new}$ discovered by traversing through the paths. In theory, if we can archive these embedded resources once, they should be available for their peer mementos while in the archives.

%%%divided by 5.5

\begin{figure}
\centering
\includegraphics[width=0.45\textwidth]{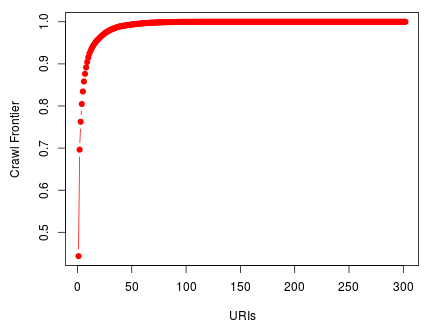}
\caption{Contributions of each URI-R and its descendants to $RP$.}
\label{cdf}
\end{figure}

The resources in Table \ref{counts} are mostly ad servers and data services such as Google Analytics. The top 300 occurring embedded resources in our entire crawl frontier are graphed -- in order of most frequent to least frequently occurring -- in Figure \ref{longtail}. Figure \ref{cdf} is a CDF measure of $R_{new}$ by URI-Rs with deferred representations and measures the deduplicated crawl frontier if we crawl all descendants for a URI-R. This shows that the largest 10\% of our frontier contributes 91\% of $RP$; that is, a large portion of the discovered crawl frontier is shared by our seed list.

%\begin{figure}
%  \begin{center}
%    \subfigure[The occurrence of embedded resources loaded into deferred representation descendants.]{\label{longtail}\includegraphics[width=0.48\textwidth]{./longTail.png}}\\
%    \subfigure[A CDF of the contributions of each URI-R and its descendants to $RP$.]{\label{cdf}\includegraphics[width=0.48\textwidth]{./cdfURIs.png}}  
%  \end{center}
%  \caption{A large number of embedded resources are shared, meaning if we archive them once we can reuse them in other mementos.}
%  \label{contribGraphs}
%\end{figure}

\section{Storing Descendants}
During its crawl, Heritrix stores the crawled representations
in Web ARChive (WARC) files \cite{warciso}; these WARCs are
indexed by the Wayback Machine and the mementos within
the WARCs are made accessible to users. Our prior research
has shown that representing descendants in the archives is
difficult and could even lead to URI-M collisions between
mementos \cite{idReps}. Jackson has discussed \cite{jacksonSopa} the challenges
with identifying descendants in the archives -- particularly
that those descendants that heavily leverage JavaScript to
change the representation without changing the URI-R may
lead to indexing and referencing challenges.

The International Internet Preservation Consortium proposed \cite{warcrevision, warcTargets}  an additional set of JSON metadata to better
represent deferred representations and descendants in
WARCs. We adapt the metadata to describe descendants
and include the interactions, state transitions, rendered content,
and interactive elements (Table \ref{fields}).

\begin{table}
\centering
\begin{tabular}{l | p{5cm}}
\textbf{Field Name} & \textbf{Data within field} \\
\hline
\hline
startedDateTime & Timestamp of interactions (no change from WARC Spec)\\
\hline
id & ID of $s_n$ represented by these interactions and resulting $R_n$.\\
\hline
title & URI-R of the descendant\\
\hline
pageTimings & Script of interactions (as CSV) to reach $s_n$ from $s_0$. E.g., click button A, click button B, then double click image C.\\
\hline
comment & Additional information \\
\hline
renderedContent & The resulting DOM of $s_n$.\\
\hline
renderedElements & $RP$ from $s_0$ to $s_n$.\\
\hline
map & The set of interactions available from $s_n$ that will transition to a new $s_{n+1}$.\\
\hline
\end{tabular}
  \caption{JSON object representing $s_n$ stored as the metadata of a WARC.}
  \label{fields}
\end{table}

%\begin{figure}
%\begin{Verbatim}
%"pages": [
%  {
%    "startedDateTime": "2015-04-20T01:00:00.000",
%    "id": "s_1_testPage.html",
%    "title": "Test Page",
%    "pageTimings": {null},
%    "comment": "state 1 of the test page",
%    "renderedContent": {<html>... <\html>},
%    "renderedElements": ["./20140907_090111.jpg",
%        "./20140903_200818.jpg"],
%    "map": [
%        {
%            "href": "img1",
%            "location": {...}
%        },
%        {
%        "href": "img2",
%        "location": {...}
%        },
%    ]
%  }
%]
%\end{Verbatim}
%\caption{JSON MetaData to be added to WARCS.}
%  \label{jsonMeta}
%\end{figure}

We present a summary of the storage requirements for descendants (using \emph{Max Coverage}) in Table \ref{storages}. If we write out
the JSON describing the states, transitions, rendered content,
and other information, it would add, on average, 16.45
KB per descendant or memento. With 8,691 descendants, a
total of 143 MB of storage space for the WARCs will be required
just for the metadata, along with the storage space for
the representations of the 54,378 new embedded resources.

The embedded resources discovered in our crawl average
2.5 KB in size. The embedded resources at $s_0$ were 2.6 KB
on average, and the newly discovered embedded resources,
as a result of deferred representations, were 2.4 KB in size
on average. We estimate that nondeferred representations,
which have 31.02 embedded resources on average, would require
80.7 KB per URI-R, or 11.1 MB of storage for the 137
URI-Rs in the collection. The storage requirement increases
to 13.4 MB with the additional metadata.

Deferred representations have 25.4 embedded resources at
$s_0$, or 70.0 KB per URI-R. For the 303 URI-Rs in the collection,
$s_0$ would require 21.2 MB of storage. In $s_1$, the crawl
discovered 45,015 embedded resources which requires 108.0
MB of additional storage, and 9,363 embedded resources
at $s_2$, or an additional 22.5 MB of storage. In all, the
303 deferred representations require 156.7 MB of storage
for the entire collection and all crawl levels (8,691 descendants).
This is 11.3 times more storage than is required for
the nondeferred representations, or 5.12 times more storage
per URI-R crawled.

If we consider the July 2015 Common Crawl \cite{iaSize} as representative
of what an archive might be able to crawl in one
month (145 TB of data for 1.81 billion URIs), an archive
would require 597.4 TB of additional storage for descendants
(29.9 TB of which is additional storage for metadata)
for a total of 742.4 TB to store descendants and metadata
for a one-month crawl. If we assume that the July crawl is a representative monthly sample, an archive would need 8.9 PB of storage for a year-long crawl.  This is an increase of 7.17 PB per year (including 358.8 TB of storage for metadata) to store the resources from deferred representations. 
Alternatively, can also say that an archive
will miss 6.81 PB of embedded resources per year because
of deferred representations.

\begin{table}
\centering
\begin{tabular}{p{5cm} | p{2cm}}
\textbf{Storage Target} & \textbf{Size} \\
\hline
\hline
JSON Metadata per descendant/memento & 16.5 KB \\
JSON Metadata of all descendants & 143 MB \\
\hline 
\multicolumn{2}{c}{Nondeferred (137 URIs)}\\
\hline
Average Embedded Resource & 2.5 KB\\
Embedded Resources per URI & 80.7 \\
Total embedded resource storage & 11.1 MB\\
Total JSON MetaData storage & 2.3 MB\\
Total with JSON Metadata & 13.4 MB\\
\hline 
\multicolumn{2}{c}{Deferred (303 URIs)}\\
\hline
Average Embedded Resource & 2.6 KB\\
Embedded Resources per URI & 70.0 \\
Embedded resource storage $s_0$ & 21.2 MB\\
Embedded resource storage $s_1$ & 108.0 MB\\
Embedded resource storage $s_2$ & 22.5 MB\\
Total JSON Metadata Storage & 5.0 MB \\
Total with JSON Metadata & 156.7 MB \\
\hline
\end{tabular}
  \caption{The storage impact of deferred representations and their descendants is 5.12 times higher per URI-R than archiving nondeferred representations.}
  \label{storages}
\end{table}

\section{Conclusions}
In this paper, we present a model for crawling deferred representations by identifying interactive portions of pages and discovering descendants. We adapt prior work by Dincturk et al. and present a FSM to describe descendants and propose a WARC storage model for the descendants.

We show that, despite high standard deviations in our sample, deferred representations have 38.5 descendants per URI-R, and that deferred representations are surprisingly shallow, only reaching a depth of 2 levels. This means that deferred representations are shallower than originally anticipated (but also very broad) and therefore it is more feasible to completely archive deferred representations using automated methods than previously thought. Archives that do not execute JavaScript during archiving are incomplete; 69\% of URIs have descendants and 96\% of the embedded resources in those descendants are not archived.

Crawling all descendants (which we defined as the \emph{Max Coverage} policy) is 38.9 times slower than crawling with only Heritrix, but adds 15.60 times more URI-Rs to the crawl frontier than Heritrix alone. Using the \emph{Max ROI} policy is 27.04 times slower than Heritrix, but adds 13.40 times more URI-Rs to the crawl frontier than Heritrix alone. 
We do not recommend one policy over the other since the policy selection depends on the archival goals of coverage or speed. However, both will help increase the ability of the crawlers to archive deferred representations. 

Most of $R_{new}$ (newly discovered by traversing the paths) are unarchived (92\% unarchived at $s_1$ and 96\% at $s_2$). However, 22.4\% of the newly discovered URI-Rs match one of the top 10 occurring URI-Rs, indicating a high amount of overlap within $RP$; mostly, these are ad servers and data-services like Google Analytics. 

We will work to incorporate PhantomJS into a web-scale crawler to measure the actual benefits and increased archival coverage realized when crawling deferred representations. Because a large portion of the embedded resources we discovered originated at data services (e.g., Google Analytics, Table \ref{counts}), we will investigate the importance (using our algorithm for memento damage \cite{brunelleDamage, damageIJDL}) of the new embedded resources in $s_1$ and $s_2$. We will also work to develop an approach to solve our current edge cases (Section \ref{edge}), including a way to handle applications like mapping applications using our automated approach along with an approach using ``canned interactions''. Our goal is to understand how many executions of canned interactions are necessary to uncover an acceptable threshold of embedded resources (e.g., how many pans and zooms are needed to get all Google Maps tiles for all of Norfolk, VA, USA?). We will also investigate filling out forms similar to Rosenthal et al. \cite{dshrDlib}.

Our work presented in this paper establishes an understanding of how much web archives are missing by not accurately crawling deferred representations and presents a process for better archiving descendants. We demonstrate that archiving deferred representation is a less daunting task with regards to crawl time than previously thought, with fewer levels of interactions required to discover all descendants. The increased frontier size and associated metadata will introduce storage challenges with deferred representations requiring 5.12 times more storage.

%ACKNOWLEDGMENTS are optional
\section{Acknowledgments}
This work was supported in part by NSF 1526700 and NEH HK-50181.

%\bibliographystyle{abbrv}
%\bibliography{_mybibtex}  

\end{document}